# Phase Competition and Rutile Phase Stabilization of Growing GeO$_2$ Films by MOCVD

Imteaz Rahaman[1], Botong Li[1], Hunter D. Ellis[1], Kathy Anderson[2], Feng Liu[3], Michael A. Scarpulla[1,3], and Kai Fu[1,*]

[1]Department of Electrical and Computer Engineering, The University of Utah, Salt Lake City, UT 84112, USA

[2]Utah Nanofab, Price College of Engineering, The University of Utah, Salt Lake City, UT 84112, USA

[3]Department of Materials Science and Engineering, The University of Utah, Salt Lake City, UT 84112, USA

**Email:** kai.fu@utah.edu

**Abstract**

Rutile germanium dioxide (r-GeO$_2$) is an ultra-wide bandgap semiconductor with potential for ambipolar doping, making it a promising candidate for next-generation power electronics and optoelectronics. Growth of phase-pure r-GeO$_2$ films by vapor phase techniques like metalorganic chemical vapor deposition (MOCVD) is challenging because of polymorphic competition from amorphous and quartz GeO$_2$. Here, we introduce seed-driven stepwise crystallization (SDSC) as a segmented growth strategy for obtaining r-GeO$_2$ films on r-TiO$_2$ (001) substrate. SDSC divides the growth into repeated cycles of film deposition and cooling-heating ramps, which suppress the non-rutile phases. We demonstrate continuous, phase-pure, partially epitaxial r-GeO$_2$ (001) films of thickness ~2 μm exhibiting x-ray rocking curves with a full-width at half-maximum of ~597 arcsec. We discuss the underlying mechanisms of phase selection during SDSC growth. SDSC-based growth provides a generalizable pathway for selective vapor-phase growth of metastable or unstable phases, offering new opportunities for phase-selective thin-film engineering.

---

[*] Author to whom correspondence should be addressed. Electronic email: kai.fu@utah.edu

**1. Introduction:** Ultrawide-bandgap semiconductors such as β-Ga$_2$O$_3$, AlGaN, and diamond exhibit exceptional breakdown strength and power-handling capabilities, positioning them as key materials for high-performance applications, including power transistors, ultraviolet photodetectors, and gas sensors.[1–4] However, each material has intrinsic limitations: β-Ga$_2$O$_3$ suffers from poor thermal conductivity and a lack of p-type doping[5–8]. AlGaN exhibits reduced hole mobility, poor acceptor activation at high Al content, difficulties in n-type doping, and challenges in wafer-scale deposition[9,10], diamond also faces challenges in wafer-scale production and doping.[11,12] Rutile GeO$_2$ (r-GeO$_2$) has emerged as a promising ultrawide-bandgap (UWBG) semiconductor, offering a bandgap of ~4.4–5.1 eV[13–16] and high electron mobilities of 244 cm²/Vs ($\perp \vec{C}$) and 377 cm²/V·s (($\parallel \vec{C}$)[17]. Galazka *et al.* reported the hall electron mobility of Sb-doped r-GeO$_2$ crystals in a range of 35-50 cm²/V·s [18]. Its Baliga figure of merit 139.68 GW/cm$^{-2}$ [19] and thermal conductivity (37 and 58 W/m·K along a- and c-axes, respectively)[20] surpass those of β-Ga$_2$O$_3$, positioning it as a strong candidate for high-power applications. Theoretical studies predict ambipolar doping is possible, with calculated hole mobilities up to 27 cm$^2$/Vs ($\perp \vec{C}$) and 29 cm$^2$/Vs ($\parallel \vec{C}$)],[14,17,21]. It is this ability to dope both n-type and p-type that enables the formation of p-n junctions. Unlike water-soluble amorphous and α-quartz GeO$_2$,[22,23] the rutile phase is stable in aqueous environments,[23] and its structural compatibility with rutile oxides (e.g., SnO$_2$, TiO$_2$, SiO$_2$) offers potential for bandgap engineering across 3.03–8.67 eV,[24–28] similar to III-nitrides and (Al$_x$Ga$_{1-x}$)$_2$O$_3$.

At room temperature and 1 atm pressure, r-GeO$_2$ is believed to be the ground state or stable polymorph of GeO$_2$. At 1 atm, the α-quartz GeO$_2$ becomes more stable at 1008 °C, while r-GeO$_2$ is more stable at higher pressures. The energy differences between r- and α-quartz GeO$_2$ are rather small, while kinetic barriers for transition between the crystalline phases are rather high; this results in α-quartz GeO$_2$ being easily maintained as a metastable allotrope. Rapid growth and low temperatures can also result in metastable amorphous GeO$_2$. To orient the discussion: our MOCVD growth is typically at 925 °C to ensure cracking of the Ge precursors, which is only 85 °C lower than the transition temperature at 1 atm. However, the growth conditions of low vacuum should favor q-GeO$_2$ somewhat. Thus, in MOCVD growth, the Gibbs free energy differences between the phases are rather small, and 2$^{nd}$ order contributions to the total film free energy, like epitaxial strain, surface and interface energies, as well as kinetic barriers, can play large roles in determining the phases manifested. As we will discuss, SDSC growth in which the films undergo periods of cooling (which favor r-GeO$_2$) and epitaxial templating with rutile substrates helps to promote the r-GeO$_2$ phase relative to amorphous and α-quartz GeO$_2$. We note that the total Gibbs free energy of a volume of a substance is related to its equilibrium vapor pressure, and that the presence of imperfections from point defects to interface, strain, curvature, and surface energies may add to the ground state phase energy especially in thin films and nanostructures, making the vapor pressure or chemical potential dependent on these factors. It is only in the bulk, unstrained, equilibrium limit that the Gibbs energy, vapor pressure, and chemical potentials of each element take on singular values. The vapor pressure or chemical potential represents the kinetic competition between atom attachment and detachment to a volume of a phase; thus, we can relate the relative phase stability to trends in growth or decomposition of the different phases in a film.

Significant advancements have been made in synthesizing single-crystal r-GeO$_2$, demonstrating the viability of various growth methods such as flux and vapor transport-based techniques.[21,29] Despite these achievements, the resulting bulk crystals have been relatively small (limited to a few mm³), and techniques like flux growth often lead to unintentional incorporation of impurities, creating challenges in preserving high purity and crystal quality. Parallel to bulk crystal synthesis, epitaxial deposition methods have offered promising pathways for thin-film growth of r-GeO$_2$, with approaches including molecular beam epitaxy (MBE),[30] pulsed laser deposition (PLD),[31,32] metal-organic chemical vapor deposition (MOCVD),[33,34] and mist chemical vapor deposition (Mist-CVD).[23,35] In MBE studies, researchers successfully stabilized crystalline r-GeO$_2$ films using a (Sn, Ge)O$_2$ buffer layer on r-plane sapphire substrates, identifying an optimal growth temperature window of 420–475 °C with extremely slow growth rates around 10 nm/h.[30] Mist-CVD has emerged as an alternative, offering notably faster deposition rates exceeding 1 μm/h on (001) r-TiO$_2$ substrates.[35] Additionally, studies at slower growth rates (~50 nm/h) showed improved alignment along the (001) orientation; however, challenges remain regarding film crystallinity, as indicated by weaker peak intensities. PLD techniques enabled direct deposition onto m-plane and c-plane sapphire at rates around 210 nm/h[31,32], although the resulting films exhibit rotational domains and partially amorphous characteristics, highlighting the need for further refinement to achieve uniform crystalline quality. These studies highlight the critical importance of phase control, as phase purity and crystalline orientation directly impact electronic properties in oxide devices. For example, phase engineering in SnO$_2$, VO$_2$, and TiO$_2$ thin films has enabled enhanced carrier mobility, optical transparency, and switching performance in transparent electronics and memristors, demonstrating the transformative role of phase-selective growth in functional oxide materials [36–39]. MOCVD is one of the most popular tools for the mass production of growing semiconductor materials. We have recently reported the first demonstration of single-crystalline r-GeO$_2$ films on r-TiO$_2$ substrates using MOCVD, exhibiting a full-width half maximum (FWHM) of ~650 arcsec.[19] To address the phase competition, the SDSC method was recently introduced, enabling improved rutile phase coverage and crystallinity.[33] However, the detailed evolution of phases—including rutile, quartz, and amorphous—during the stepwise growth process requires further investigation to elucidate the underlying thermodynamics, kinetic pathways, and phase transition mechanisms governing the MOCVD deposition of $GeO_2$.

In this study, the SDSC with a new scheme has been employed to study the phase competition and transitions. By systematically varying the segmentation intervals during deposition while keeping the total growth period constant, we have explored how controlled interruptions enhance rutile phase formation during the phase competition with quartz and amorphous $GeO_2$ and improve both the crystalline coverage and stabilization of the rutile phase. Comprehensive characterization techniques—including high-resolution X-ray diffraction (HR-XRD), scanning electron microscopy (SEM), atomic force microscopy (AFM), and high-resolution transmission electron microscopy (HR-TEM)—were employed to evaluate the crystal quality, surface morphology, microstructure, and interfacial characteristics of the films. This work provides new insights into the crystallization pathways of $GeO_2$ and establishes fundamental guidelines for growing phase-pure and high-quality rutile $GeO_2$ films via MOCVD.

## 2. Experimental details

Unintentionally doped $GeO_2$ thin films were deposited using an MOCVD system from Agnitron Technology. All growth segments were performed at 925 °C under a constant total chamber pressure of 80 Torr. Tetraethyl germane (TEGe) served as the germanium source, while high-purity $O_2$ acted as the oxidizing agent. Argon (Ar) was employed both as the carrier and the shroud gas. The precursor flow rates were precisely controlled at $1.35 \times 10^{-5}$ mol/min for TEGe and $8.94 \times 10^{-2}$ mol/min for $O_2$, and the susceptor was rotated at 300 RPM to promote uniform film deposition across the substrate surface.

Rutile (001) TiO$_2$ is selected as the substrate due to its small lattice mismatch (~4.1%) with rutile-phase GeO$_2$, providing a favorable template for epitaxial growth. However, both the amorphous and quartz phases of GeO$_2$ possess higher entropic contributions and lower Helmholtz free energy compared to the rutile phase under typical MOCVD growth conditions.[40] This thermodynamic advantage, particularly at elevated temperatures and during prolonged deposition, drives a competitive evolution toward the quartz and/or amorphous phases over film thickness. As a result, phase competition among amorphous, quartz, and rutile GeO$_2$ emerges during vapor-phase growth, critically influencing the final film structure. Nevertheless, under the sub-atmospheric pressure typically used in MOCVD, quartz may gain relative thermodynamic favorability, as lower pressures tend to stabilize the quartz phase. The close proximity in free energy among the rutile, quartz, and amorphous phases under these conditions likely contributes to their concurrent presence during growth. While quartz is metastable at the growth temperature, its kinetic accessibility facilitates its formation alongside amorphous regions [41]. However, during cool-down or re-heating—particularly in the case of growth interruptions, these less stable phases may partially decompose and desorb due to their higher volatility and lower thermal stability compared to rutile, rather than transforming into a more stable crystalline phase.

To explore phase competition and transition mechanisms during growth, compared to conventional continuous growth [**Figure 1a**], four distinct strategies were applied using the SDSC segmented growth [**Figure 1b**]—multiple shorter growth intervals separated by cooling/heating cycles, each under identical growth conditions (e.g., 925 °C as the growth temperature) and a fixed total duration of 180 minutes **[Figure 1c]:** a continuous 180-minute deposition used as the baseline reference (Sample A: "180 min × 1"); a 180-minute deposition but segmented into two 90-minute depositions without taking the sample out of the reactor (Sample B: "90 min × 2"); a 180-minute deposition but segmented into three 60-minute depositions without taking the sample out of the reactor (Sample C: "60 min × 3"); and a 180-minute deposition but segmented into six 30-minute depositions (Sample D: "30 min × 6"). Prior to the film growth, rutile-phase $TiO_2$ (001) substrates were subjected to a multi-step cleaning process, including piranha solution (a 3:1 mixture of $H_2SO_4$:$H_2O_2$) to remove organic residues, followed by sequential rinses in acetone, isopropanol, and deionized water to eliminate any remaining surface contaminants. Structural properties of the deposited films were characterized using high-resolution X-ray diffraction (HR-XRD) with a Bruker D8 DISCOVER system equipped with a Cu K$\alpha_1$ source ($\lambda$ = 1.5406 Å), a triple-axis channel-cut monochromator, and an Eiger R 250K area detector. Surface topography and morphology were examined using a Bruker Dimension ICON atomic force microscope (AFM) and a Quanta 600F environmental scanning electron microscope (SEM) equipped with microscale EDS capabilities.

## 3. Results and analysis:

The schematic in Figure **1c** summarizes the segmented growth strategies, showing that while the total growth time is kept constant (180 min), the division into 1, 2, 3, or 6 segments leads to distinct outcomes. SEM images in Figure **1d** show that continuous 180 min growth yields sparse rutile grains on an amorphous matrix, whereas the corresponding AFM image in Figure **1e** reveals a high surface roughness (RMS ≈ 117.3 nm). With two 90-minute segments, Figure **1f** shows the coexistence of rutile and quartz domains, and Figure **1g** captures a slight reduction in roughness (RMS ≈ 83.1 nm). Further segmentation into three 60-minute cycles (Figure **1h**) enhances quartz and rutile formation, with Figure **1i** showing a marked reduction in roughness (RMS ≈ 20.5 nm). Finally, the 30 min × 6 SDSC condition (Figure **1j**) produces continuous, faceted rutile coverage, as further evidenced by the AFM image in Figure **1k**, which exhibits a moderate roughness increase (RMS ≈ 77.8 nm) due to its faceted crystalline texture. Collectively, these results validate SDSC as an effective pathway engineering technique that selectively amplifies rutile crystallization by harnessing thermodynamic, kinetic, and interfacial control mechanisms.

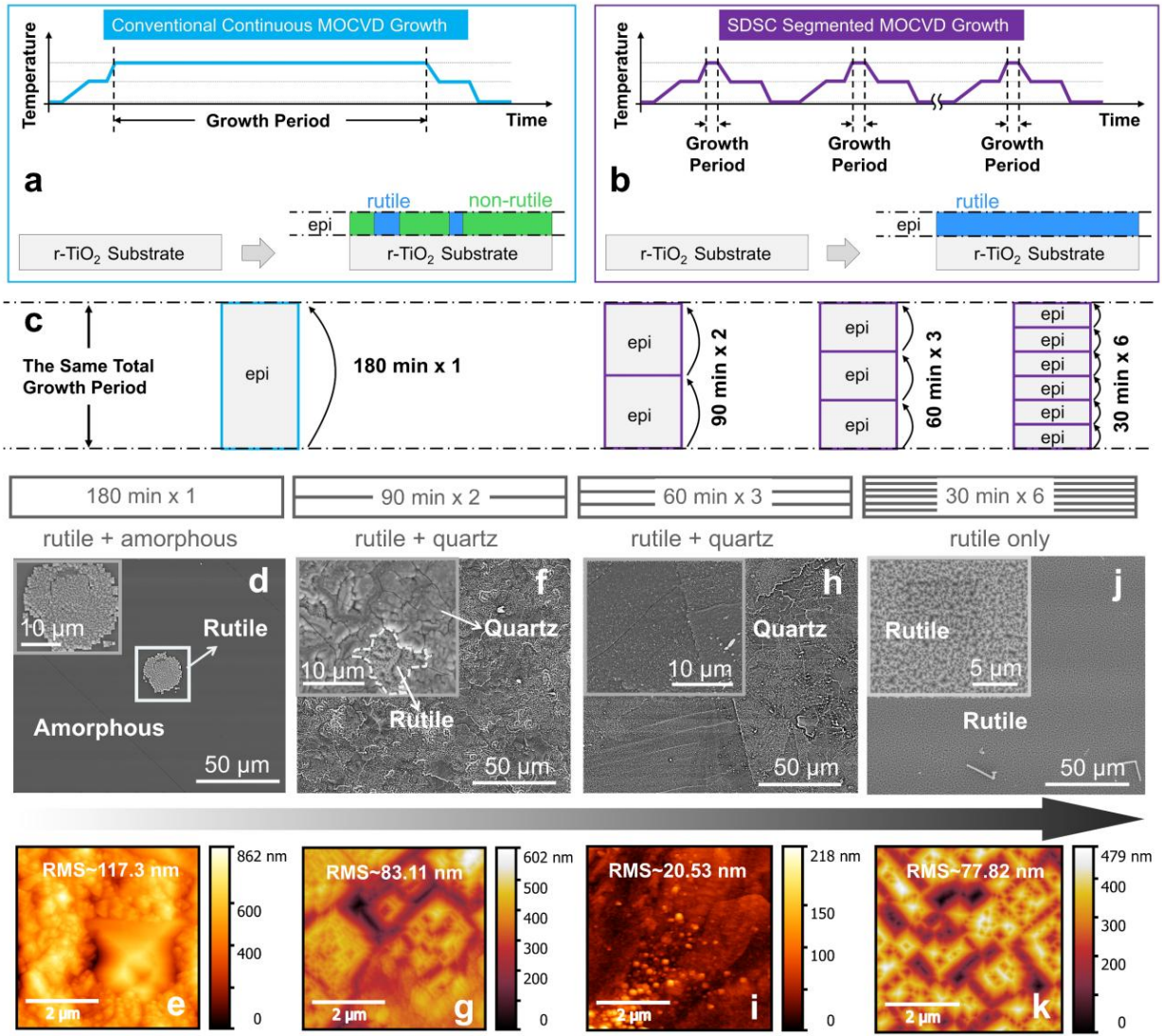

**Figure 1.** Growth behavior and morphology of GeO$_2$ films on r-TiO$_2$ (001) using continuous and segmented MOCVD. (a) Schematic of continuous growth (180 min × 1), leading to dominant amorphous and quartz phases. (b) SDSC strategy with segmented growth and intermediate cooling, promoting rutile formation. (c) Comparison of samples from different growth schemes (180 min

× 1, 90 min × 2, 60 min × 3, and 30 min × 6). (d–k) SEM (top) and AFM (bottom) images showing morphology evolution. Continuous growth yields mixed phases and high roughness (d, e), intermediate segmentation forms rutile–quartz mixtures (f–i), and SDSC achieves uniform, phase-pure rutile films with faceted grains (j, k).

**Figure 2** provides an evolution of the nucleation dynamics and phase competition of $GeO_2$ during continuous MOCVD growth, complementing the phase evolution described in **Figure 1**. The XRD patterns in **Figure 2a** capture the structural progression of the films with growth durations ranging from 1 to 90 minutes. In the early stages (1–30 min), the absence of discernible diffraction peaks suggests the formation of amorphous material or extremely fine nuclei below the detection limit, consistent with the SEM images in **Figures 2b–f** that display largely smooth surfaces with minimal signs of crystallization. As growth continues for 60 minutes, weak peaks corresponding to emerging $GeO_2$ phases appear (**Figure 2a**), accompanied by scattered island nucleation observed in **Figure 2g**. By 90 minutes, a weak rutile (r-$GeO_2$) (002) peak becomes visible, and SEM imaging (**Figure 2h**) confirms the presence of isolated rutile islands coexisting with an amorphous matrix. This delayed and spatially sparse rutile formation reflects sluggish nucleation kinetics and limited surface coverage under continuous growth conditions. The gradual increase in peak intensity and sharpening with time signifies both vertical and lateral growth of rutile domains in the later stages, though the process remains constrained by persistent phase competition. This trend reinforces the hypothesis in **Figure 1** that continuous deposition promotes competing nucleation pathways where quartz and amorphous phases tend to dominate over rutile. Such competition is further visualized in **Figures 2i–p**, which reveal a wide variety of surface morphologies after a single uninterrupted 60-minute growth. The observed features include dispersed nanocrystalline patterns (Nanodot Constellation, **Figure 2i**), compact aggregated clusters (Rosette Clusters, **Figure 2j**), extended dendritic structures (Fern-Like Branches, **Figure 2k**), and large radial domains (Spherulites, **Figure 2l**), along with more complex formations such as the irregularly shaped Fragmented Starburst (**Figure 2m**), the radially symmetric Frosted Dendritic Halo (**Figure 2n**), the interwoven Dense Coral Web (**Figure 2o**), and the fine-grained Nanograin Carpet (**Figure 2p**). These diverse morphologies underscore the simultaneous growth of rutile, quartz, and amorphous phases—each governed by differences in surface mobility, interfacial energy, and growth kinetics. In particular, the frequent emergence of quartz-related spherulitic structures suggests their competitive advantage in many regions, likely due to favorable thermodynamic and kinetic conditions that enable quartz to overgrow and obscure underlying rutile seeds. This overgrowth mechanism contributes to the buried or patchy nature of rutile domains in prolonged deposition, as also seen in **Figure 2h**. In contrast, the segmented growth approach described in **Figure 1** disrupts this phase competition by introducing periodic nucleation resets, allowing rutile grains to consistently emerge and expand without being overtaken. These results align with classical nucleation theory[42], where phase selection is dictated by the interplay between kinetic accessibility and thermodynamic stability. Overall, results in **Figure 2** confirm the phase competition during MOCVD growth of $GeO_2$.

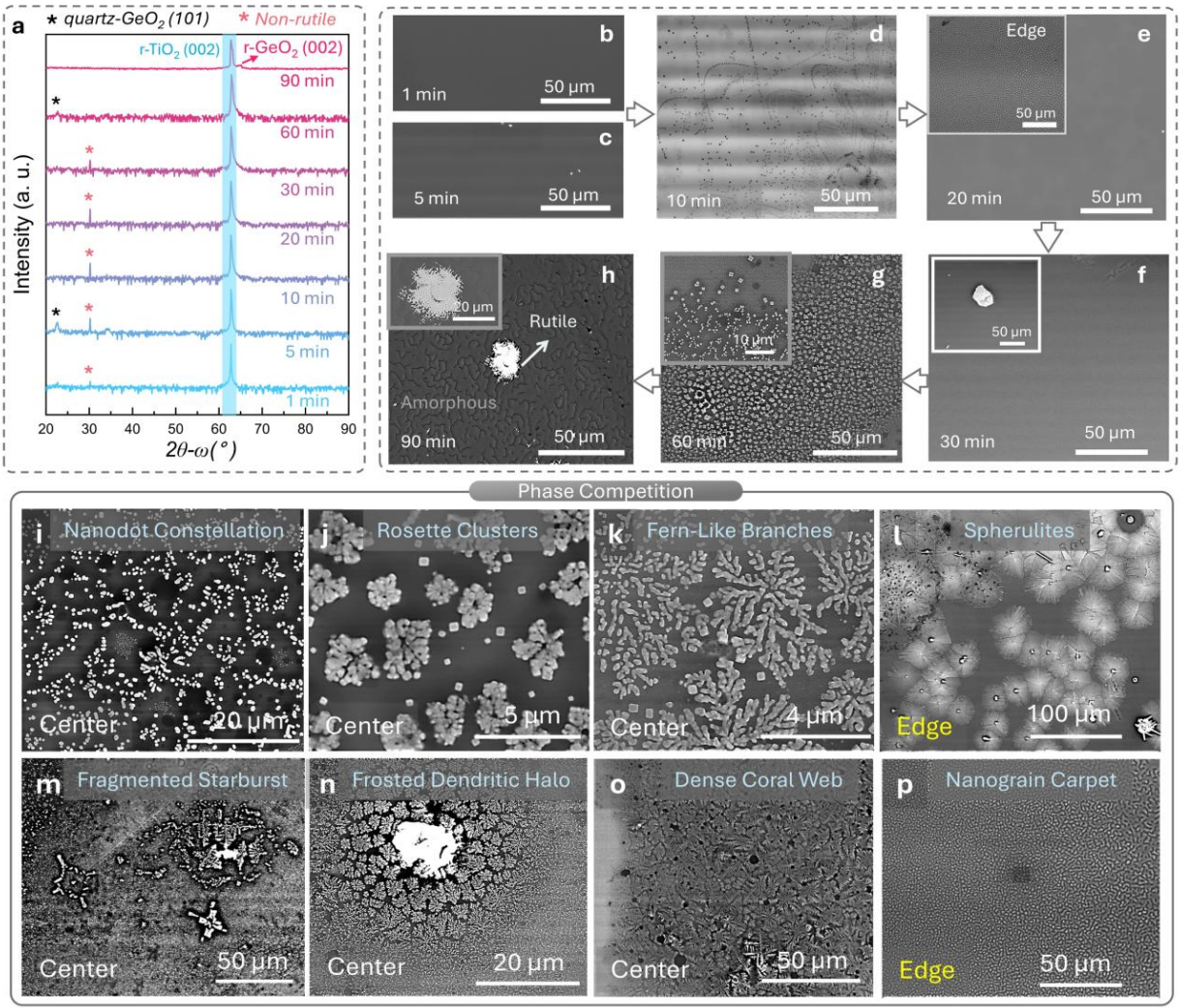

**Figure 2.** Phase evolution and morphology of GeO$_2$ films under continuous MOCVD. (a) XRD patterns show no crystalline peaks up to 30 min; rutile (002) appears weakly at 90 min. (b–h) SEM images reveal smooth surfaces up to 30 min, with island formation at 60 min and sparse rutile crystallites at 90 min. (i–p) SEM images from a 60 min film show phase competition, where quartz-like spherulites dominate and obscure underlying rutile grains.

Two key features of the SDSC segmented growth may contribute to the enhanced formation of the rutile phase: (1) segmentation of the overall growth period into shorter deposition intervals, and (2) the introduction of additional heating and cooling cycles between segments. **Figure 3** examines the role of unintentional annealing during the SDSC process and its effect on phase transformation dynamics in $GeO_2$ films. The SDSC process involves repeated short growth segments separated by thermal ramps, during which the precursor is turned off and the temperature cycles from deposition (~925 °C) down to idle (~725 °C). These intervals, although intended for segmentation, inherently introduce multiple annealing steps. To isolate the effect of this unintentional annealing, a control experiment was designed as illustrated in **Figure 3a**, applying identical thermal cycling (cooling/heating cycles) without precursor flow to simulate the annealing conditions of SDSC without deposition. SEM images before and after this annealing-only treatment (**Figure 3b and 3c**) reveal a significant morphological evolution. Initially, the film comprises distinct regions of amorphous, rutile, and quartz phases, with amorphous areas (darkest contrast) predominating. After annealing, a substantial portion of the amorphous regions transform into crystalline quartz, as indicated by the transformation from smooth surface (indicated with yellow color) in **Figure 3b** to "cracked" surface (indicated with orange color) in **Figure 3c**. In contrast, rutile islands marked in both panels remain largely unchanged, with minimal lateral growth into surrounding amorphous zones (indicated with white circles). This observation suggests that the amorphous-to-quartz transition is thermodynamically favored during passive thermal cycling, while the conversion to rutile is kinetically limited and strongly dependent on the presence of nearby rutile nuclei. The enhanced quartz formation in the absence of precursor further underscores the inherent tendency of $GeO_2$ to crystallize into quartz under prolonged sub-saturation heating, unless continuously driven by precursor supply and controlled nucleation as in SDSC. These findings emphasize that while unintentional annealing promotes phase transformation, it favors quartz over rutile in the absence of active nucleation, reaffirming the critical role of segmented precursor-supplied growth in enabling sustained rutile-phase development. The SDSC process not only resets nucleation conditions to suppress quartz dominance but also enhances the rutile phase nucleation and expansion, while rutile seeds formed in each cycle can serve as effective nucleation sites for next growth cycles, thereby eventually achieving phase-pure r-GeO$_2$.

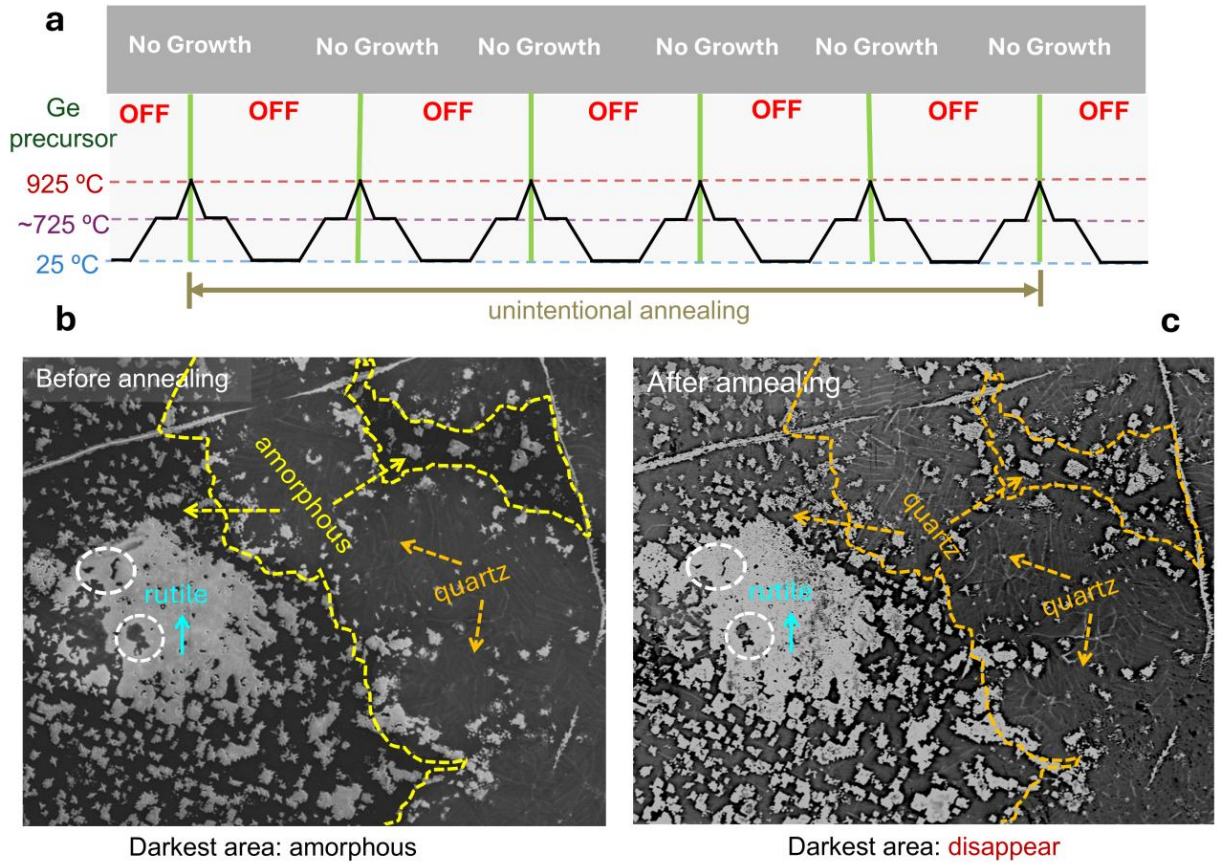

**Figure 3.** Role of unintentional annealing in SDSC growth. (a) Control annealing without precursor. (b, c) SEM images before and after annealing show that amorphous regions convert to quartz, while rutile remains unchanged, highlighting the dominance of the quartz phase in the absence of active growth.

**Figure 4** examines the influence of segmented deposition schemes on the crystalline phase evolution and out-of-plane alignment of GeO$_2$ films. As shown in **Figure 4a**, XRD *2θ–ω* scans indicate a clear trend from predominantly amorphous or quartz-rich phases in the 180 min × 1 and 90 min × 2 samples toward a more pronounced rutile (002) peak in the 30 min × 6 film, confirming the effectiveness of the SDSC strategy in promoting rutile phase formation. Intermediate schemes such as 90 min × 2 and 60 min × 3 exhibit coexisting quartz and rutile peaks, reflecting the phase competition dynamics described in **Figures 1 and 2**. Notably, the rutile signal intensifies as the deposition is further segmented, highlighting the advantage of shorter intervals in suppressing metastable phase formation. **Figure 4b** presents the rocking curves (Δω) of the rutile (002) peak for each sample, where the full width at half maximum (FWHM) reflects the degree of out-of-plane crystalline alignment. The segmented schemes all show improved crystallinity compared to the continuous 180 min × 1 film, with the 60 min × 3 sample yielding the narrowest FWHM (~0.159°), indicating excellent alignment despite partial phase competition. The 30 min × 6 films, while phase-pure and more surface-covering, exhibit a broader FWHM (~0.413°), suggesting smaller domain sizes likely due to frequent nucleation resets. To further enhance both crystallinity and surface coverage, the segmented scheme was extended to 30 min × 12 (**Figure 4c**), which shows a stronger rutile (002) reflection relative to the substrate and a minor quartz signal originating from the previously amorphous center region. Rocking curve analysis in **Figure 4d** confirms that the FWHM significantly narrows to 0.166° (597.6 arcsecs) in the 30 min × 12 sample, reflecting improved grain size and out-of-plane order. **Figure 4e** summarizes the FWHM evolution for all deposition schemes, illustrating that extended SDSC growth not only maintains phase purity but also enables nearly full surface coverage with high crystalline quality. Collectively, these results underscore that the SDSC method, particularly with extended segmented cycles, provides a powerful approach to control both phase selectivity and crystal quality, achieving continuous, phase-pure rutile $GeO_2$ films suitable for scalable device integration.

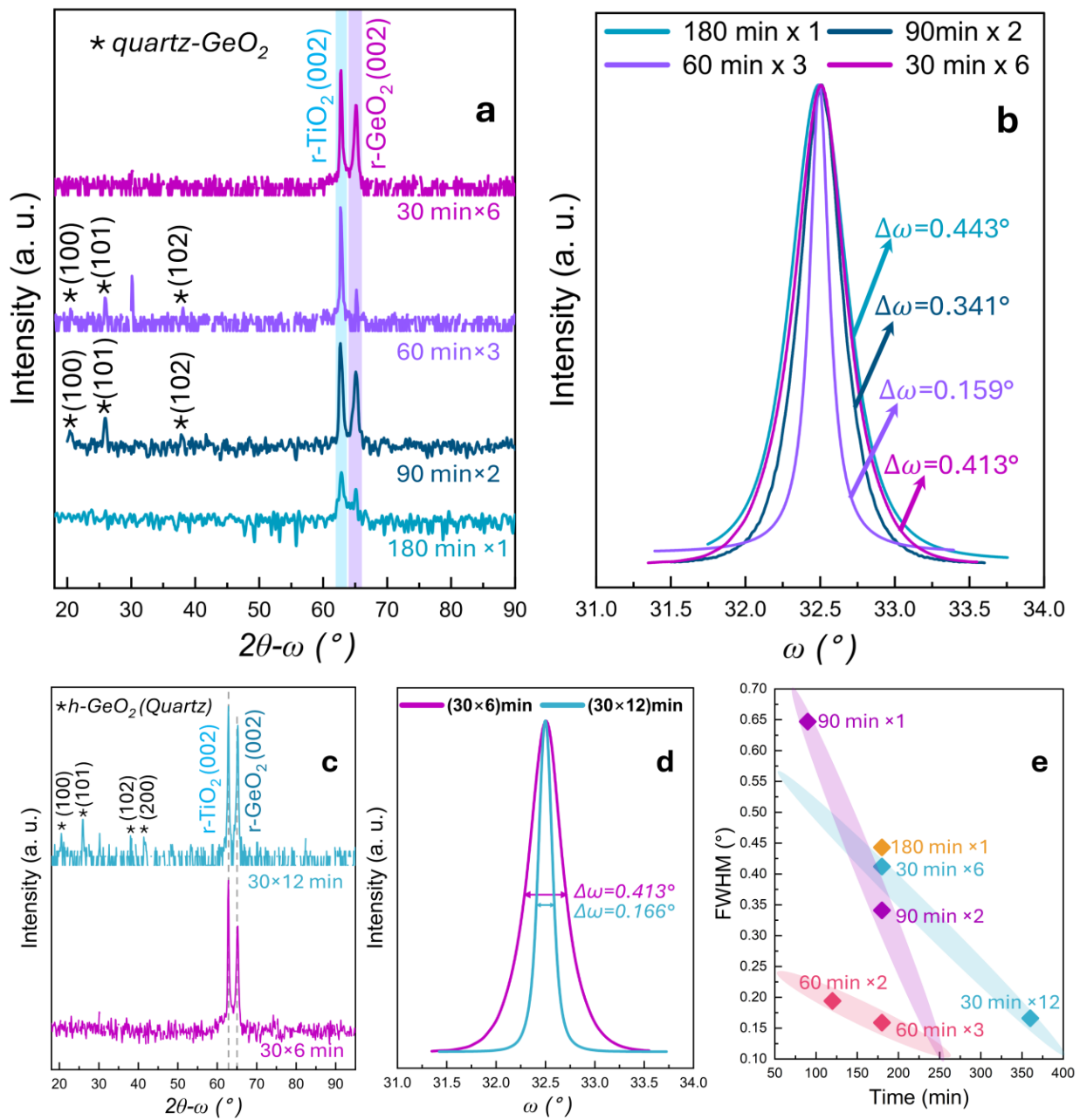

**Figure 4.** Structural evolution of GeO$_2$ films under various segmented SDSC growth schemes. (a) XRD *2θ–ω* scans show a transition from non-rutile phases in continuous "180 min × 1" to dominant rutile (002) peaks in segmented films, with increased segmentation favoring rutile formation. (b) Rocking curves (Δω) reveal improved out-of-plane crystallinity for segmented films, with the narrowest FWHM observed in the "60 min × 3" sample. (c) Comparison of "30 min

× 6" and "30 min × 12" samples show enhanced rutile intensity and the emergence of quartz in the center region with extended growth. (d) Rocking curve FWHM narrows from 0.413° to 0.166°, indicating improved grain alignment. (e) FWHM trends confirm that prolonged segmented growth yields higher-quality rutile $GeO_2$ with near-complete surface coverage.

**Figure 5** provides a detailed TEM investigation of the rutile $GeO_2$ film grown on rutile $TiO_2$ for "30 min × 12" sample. In **Figure 5a**, a low-magnification cross-sectional TEM image shows the protective Pt layer on top, the rutile $GeO_2$ film in the middle, and the $TiO_2$ substrate below; the violet rectangle highlights the region selected for higher-resolution analysis. It is worth noting that the film exhibits a thickness of approximately 2.01 μm, corresponding to a growth rate of about 340 nm/h. The elemental maps in **Figures 5b, 5c, and 5d** verify the distributions of oxygen (O), titanium (Ti), and germanium (Ge), respectively, confirming a sharp chemical boundary between $GeO_2$ and $TiO_2$. Notably, there is no discernible diffusion of Ti into the $GeO_2$ layer. At higher magnification, **Figure 5e** displays a high-resolution TEM image of rutile $GeO_2$, revealing the atomic-scale arrangement of the lattice with well-defined fringes along the [001] growth direction and the in-plane [100] axis. The corresponding inverse Fast Fourier Transform (FFT) image in **Figure 5f** highlights the measured lattice spacings, which match characteristic rutile $GeO_2$ planes. These spacings are slightly larger than theoretical values, likely due to residual stress within the film. The selected-area electron diffraction (SAED) pattern in **Figure 5g**, indexed to the [0–10] zone axis, further confirms the crystalline quality of rutile $GeO_2$. The Miller indices corresponding to each SAED spot were identified using the crystallographic analysis software CrysTBox. Finally, **Figure 5h** shows the interface region, where the boundary between $GeO_2$ and $TiO_2$ is somewhat zigzag rather than flat, possibly reflecting slight substrate decomposition during crystalline film formation. The inset FFT in this panel indicates a coherent interface with no observable amorphous interlayer, underscoring the high-quality rutile GeO$_2$ film achieved on rutile TiO$_2$.

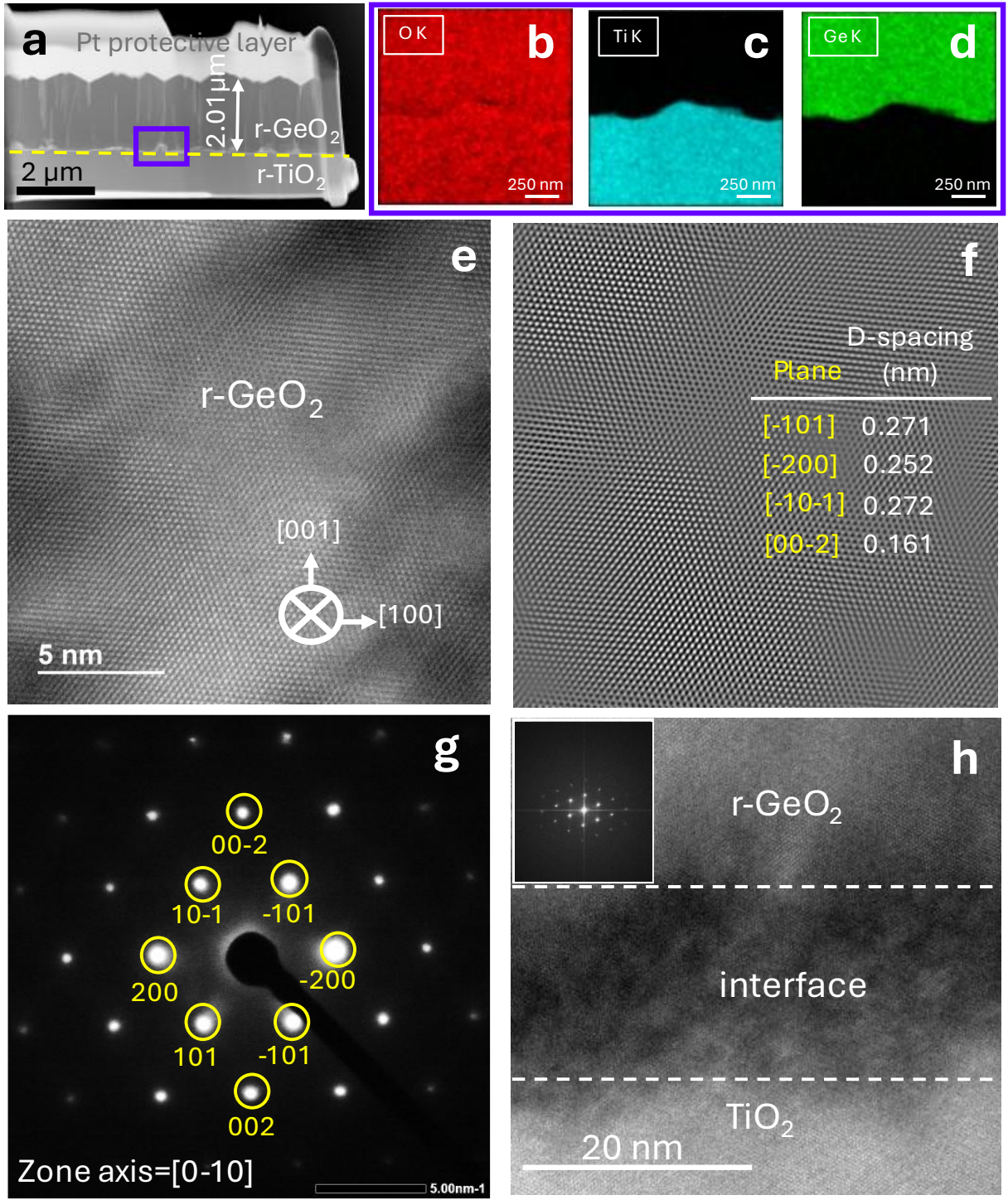

**Figure 5.** (a) Low-magnification cross-sectional TEM image of ~2.01 μm-thick r-GeO$_2$ on r-TiO$_2$. (b–d) EDS elemental maps of O, Ti, and Ge showing clear separation and no Ti diffusion. The region for (b-d) corresponds to the region marked in (a) by the violet box. (e) High-resolution TEM image with [001] out-of-plane and [100] in-plane orientation. (f) Bragg-filtered inverse FFT image highlighting lattice planes. (g) SAED pattern indexed to the [0–10] zone axis confirming high crystallinity. (h) TEM image of the GeO$_2$/TiO$_2$ interface showing a coherent zigzag boundary with no amorphous interlayer.

Based on the experimental results, we propose the following growth hypothesis:

**Hypothesis 1:** Continuous MOCVD growth promotes nucleation of amorphous and quartz GeO$_2$ phases due to thermodynamic favorability.

Evidence: Continuous 180 min deposition yields films dominated by amorphous and quartz phases, with only sparse rutile crystallites, reflecting the higher entropy and lower Helmholtz free energy of amorphous and quartz phases at typical growth conditions (925 °C, 80 Torr). The epitaxial influence of the rutile TiO$_2$ substrate is insufficient to suppress competing phase nucleation over prolonged deposition.

**Hypothesis 2:** Existing rutile seeds enable preferential lateral expansion and coalescence during subsequent deposition cycles.

Evidence: Additional MOCVD runs on mixed-phase films result in progressive rutile coverage, indicating that rutile regions provide lower interfacial energy and nucleation barriers than amorphous or quartz regions. The rutile seeds provide an ordered lattice surface, offering low-energy sites for adatoms and incoming GeO$_2$. In contrast, amorphous regions lack a structural template, and polycrystalline quartz presents higher interfacial energy and nucleation barriers due to lattice mismatch and domain boundaries.

**Hypothesis 3:** Segmented SDSC deposition with intermediate cooling resets nucleation conditions, favoring rutile growth.

Evidence: SDSC growth (e.g., 30 min × 6) consistently yields phase-pure rutile films, in contrast to continuous or longer segmented growth, which favors competing phases. Each deposition restart produces a transient local supersaturation peak, resulting in a reduced critical nucleus size that enhances nucleation rates at energetically preferred rutile seeds rather than the formation of new amorphous or quartz nuclei.

**Hypothesis 4:** Cooling-heating cycles act as kinetic barriers, suppressing amorphous-to-quartz transformation.

Evidence: Control experiments show that annealing without a precursor promotes amorphous-to-quartz transition, but segmented growth prevents this by interrupting the sustained high-temperature conditions required for such transformation. Rutile seeds remain stable during these cycles, keep the surface ready for vapor-driven rutile growth at seeds, and dominate the growth front upon restarting the next segmented growth period.

### 4. Conclusion

Unlike conventional semiconductors commercialized today—where a single thermodynamically stable phase or stacking-related polytypes typically form under MOCVD growth dictated by substrate orientation—$GeO_2$ presents a unique challenge due to its intrinsic polymorphic competition among three structurally distinct phases: rutile, quartz, and amorphous. This inherent phase instability makes the realization of phase-pure rutile $GeO_2$ films particularly difficult. To address this, a segmented deposition with seed-driven stepwise crystallization (SDSC) has been demonstrated to enable phase-pure, high-quality rutile $GeO_2$ thin films by overcoming the intrinsic polymorphic competition. Compared to continuous 180-minute deposition, which predominantly yields amorphous films, segmented deposition with 30-minute intervals (30 min × 6) resulted in a rutile-dominant film, characterized by a rutile (002) rocking curve FWHM of 0.413°. Further extension of the SDSC sequence to "30 min × 12" led to a significant enhancement in crystalline quality, evidenced by a reduced FWHM of 0.166° and nearly 100% surface coverage of the rutile phase. XRD, SEM, and AFM measurements all corroborate the improved crystallinity and phase selectivity. These findings demonstrate that the SDSC approach effectively mitigates undesired phase transitions and promotes selective stabilization of rutile $GeO_2$, offering a viable pathway to achieving high-quality, phase-pure, and continuous rutile films. Beyond $GeO_2$, this strategy opens up new possibilities for tailoring polymorphic systems via kinetic control in vapor-phase deposition.

## AUTHOR DECLARATIONS

### Conflict of Interest

The authors have no conflicts to disclose.

### Author Contributions

**Imteaz Rahaman:** Data curation (lead); Formal analysis (lead); Investigation (lead); Methodology (lead), Writing-original draft (lead). **Botong Li:** Writing – review & editing (supporting). **Hunter D. Ellis:** Writing – review & editing (supporting); **Kathy Anderson:** Investigation (supporting). **Feng Liu:** Formal analysis (Supporting); Writing – review & editing (supporting). **Michael A. Scarpulla:** Conceptualization (Supporting); Formal analysis (Supporting); Writing – review & editing (supporting). **Kai Fu:** Conceptualization (lead);

Supervision (lead); Project administration (lead); Resources (lead), Writing – review & editing (lead).

## ACKNOWLEDGEMENT

This work made use of Nanofab EMSAL shared facilities of the Micron Technology Foundation Inc. Microscopy Suite sponsored by the John and Marcia Price College of Engineering, Health Sciences Center, Office of the Vice President for Research. In addition, it utilized the University of Utah Nanofab shared facilities, which are supported in part by the MRSEC Program of the NSF under Award No. DMR-112125.

## DATA AVAILABILITY

The data that supports the findings of this study are available from the corresponding authors upon reasonable request.